\documentclass[twocolumn,preprintnumbers,amsmath,amssymb,floatfix, superscriptaddress]{revtex4-2}

\usepackage{graphicx}
\usepackage{dcolumn}
\usepackage{bm,epsfig}
\usepackage{epstopdf}
\usepackage{xcolor}
\usepackage[percent]{overpic}

\begin{document}

\title{Effects of external pressure on the narrow gap semiconductor Ce$_{3}$Cd$_{2}$As$_{6}$}

\author{M. M. Piva}
\email{Mario.Piva@cpfs.mpg.de}
\affiliation{Instituto de F\'{\i}sica ``Gleb Wataghin'', UNICAMP, 13083-859, Campinas, SP, Brazil}
\affiliation{Los Alamos National Laboratory, Los Alamos, New Mexico 87545, USA}
\affiliation{Max Planck Institute for Chemical Physics of Solids, N\"{o}thnitzer Str.\ 40, D-01187 Dresden, Germany}

\author{L. Xiang}
\affiliation{Ames Laboratory, Iowa State University, Ames, Iowa 50011, USA}
\affiliation{Department of Physics and Astronomy, Iowa State University, Ames, Iowa 50011, USA}

\author{J. D. Thompson}
\affiliation{Los Alamos National Laboratory, Los Alamos, New Mexico 87545, USA}

\author{S. L. Bud'ko}
\affiliation{Ames Laboratory, Iowa State University, Ames, Iowa 50011, USA}
\affiliation{Department of Physics and Astronomy, Iowa State University, Ames, Iowa 50011, USA}

\author{R. A. Ribeiro}
\affiliation{Ames Laboratory, Iowa State University, Ames, Iowa 50011, USA}
\affiliation{Department of Physics and Astronomy, Iowa State University, Ames, Iowa 50011, USA}

\author{P. C. Canfield}
\affiliation{Ames Laboratory, Iowa State University, Ames, Iowa 50011, USA}
\affiliation{Department of Physics and Astronomy, Iowa State University, Ames, Iowa 50011, USA}

\author{P. F. S. Rosa}
\affiliation{Los Alamos National Laboratory, Los Alamos, New Mexico 87545, USA}

\date{\today}

\begin{abstract}

Here we report the magnetic and electronic properties of recently discovered Ce$_{3}$Cd$_{2}$As$_{6}$. At ambient pressure, Ce$_{3}$Cd$_{2}$As$_{6}$ presents a semiconducting behavior with an activation gap of 74(1)~meV. At 136~K, a sudden increase of the electrical resistivity and a peak in specific heat are consistent with a charge density wave transition. At low temperatures, antiferromagnetic order of the Ce$^{3+}$ ions occurs below $T_{\rm N} = 4.0$~K with a magnetic hard axis along the $c$-axis and a $\Gamma_{6} = |\pm1/2\rangle$ doublet ground state. The application of external pressure strongly suppresses the charge density wave order, which is completely suppressed above 0.8(1)~GPa, and induces a metallic ground state. No evidence for superconductivity is detected above 2~K. Conversely, the antiferromagnetic state is favored by pressure, reaching a transition temperature of 5.3~K at 3.8(1)~GPa. Notably, the resistivity anomaly characterizing the antiferromagnetic order changes with increasing pressure, indicating that two different magnetic phases might be present in Ce$_{3}$Cd$_{2}$As$_{6}$ under pressure. This change in ordering appears to be associated to the crossing of the $T_{\rm CDW}$ and $T_{\rm N}$ lines. 

\end{abstract}

\maketitle

\section{INTRODUCTION}

Emergent phenomena in strongly correlated materials often stem from competing interactions near an order-disorder or a metal-insulator boundary \cite{StronglyCorrelated}.
A demonstrated strategy to uncover emergent phenomena is to tune new materials through these boundaries $via$ chemical doping, magnetic fields, or the application of external pressure. 
The latter is a particularly important parameter that compresses the material and tunes the electronic structure (e.g. bandwidth, hybridization, and valence) without introducing disorder \cite{McMahan,Nicklas}. 

Notably, materials containing layers of square nets provide a platform for realizing pressure-induced phenomena.
At ambient pressure, the low-dimensionality of the square nets makes them susceptible to distortions that may result in a modulation of the conduction electrons, the opening of a semiconducting gap, and the emergence of a charge density wave (CDW) \cite{CDW,CDWlowD2,ReviewCDW,ReviewCDW2,Sqrnets_1,Sqrnets_2}.
The response of CDWs under external applied pressure is often striking but not yet fully understood. On one hand, the application of pressure may increase the electron-phonon interaction, which favors the formation of CDWs \cite{ReviewCDW,4HbTaS2}. On the other hand, applied pressure might enhance the interlayer interactions, resulting in a more three-dimensional Fermi surface that suppresses the nesting vectors responsible for the CDWs \cite{4HbTaS2}. The latter effect is much pursued, as the suppression of CDWs typically favor the appearance of conventional superconductivity, due to the increase of the density of states at the Fermi level once the gap is closed and the electron-phonon interaction present in these materials, as observed in NbSe$_{3}$, Sr$_{3}$Ir$_{4}$Sn$_{13}$, Ca$_{3}$Ir$_{4}$Sn$_{13}$ and LaPt$_{2}$Si$_{2}$ \cite{CDW/SDW/SC-Peierls,CDWSCSrCa3413,LaPt2Si2}. However, unconventional superconductivity may also emerge upon suppression of a CDW phase similar to CsV$_{3}$Sb$_{5}$ \cite{CsVSb}.          

In addition, intercalating rare-earth elements with square nets 
may induce further instabilities, such as magnetism in the localized $4f$ limit
or a heavy Fermi liquid when hybridization between 4$f$ and 
square net $p$ bands is in the strong coupling limit. Once more, emergent phenomena are expected
at the localized-delocalized boundary, unconventional superconductivity being the classic example. 
In particular, several Ce-based compounds display emergent ground states and complex phase diagrams at ambient pressure and under pressure \cite{reviewHF}. Finally, the combination of strong electronic correlations 
and a narrow-gap semiconducting state is a promising route to improve thermoelectricity \cite{Thermo,CeThermo}.

The large family of ``112" materials with general formula 
Ce$M_{1-\delta}X_{2}$ ($M =$ transition or post-transition metal, $X =$ pnictide) crystallizes in the tetragonal $P4/nmm$ or $I4/mmm$ structures, which contain layers of pnictide square nets, and hosts many examples of emergent ground states. CeAgBi$_{2}$, for instance, is antiferromagnetic below 6.4~K and presents several metamagnetic transitions and an anomalous Hall contribution for fields applied parallel to $c$-axis \cite{SMT}. Moreover, CeAgSb$_{2}$ presents a ferromagnetic order below 9.8~K, in which a topological magnetic hysteresis of the domains was recently observed \cite{CeAgSb2_1}, also it presents a complex evolution of the magnetic phase as a function of applied pressure \cite{CeAgSb2_2}. CeAuSb$_{2}$ is another example of a complex magnetic structure under the application of magnetic fields \cite{Phase, Stress, Multiq, SSeo} and has been recently proposed to host a nematic state coupled to its antiferromagnetism \cite{SSeo2}. CeAuBi$_{2}$ also presents several metamagnetic transitions when fields are applied parallel to the $c$-axis at 2~K \cite{MarioAu}, which suggests a complex evolution of its magnetic structure with field. Furthermore, this family of materials is well described by a trend in the ground state wavefunction of the localized moment. For members crystallizing in the $P4/nmm$ structure, antiferromagnetism is favored by a  $\Gamma_{7}$ ground state, while ferromagnetism is observed in compounds with a $\Gamma_{6}$ ground state \cite{RosaCeCdSb}. 

Here we focus on the newly-discovered narrow-gap semiconductor Ce$_{3}$Cd$_{2}$As$_{6}$ \cite{LaCe326}, whose crystal structure
can be viewed as the ordered version of a tetragonal CeCd$_{0.67}$As$_{2}$ (112) parent subcell. In this case, the 
Cd vacancies order within the body-centered structure and give rise to the superstructure Ce$_{3}$Cd$_{2}$As$_{6}$ within a monoclinic $C2/m$ space group. At ambient pressure, Ce$_{3}$Cd$_{2}$As$_{6}$ and its non-magnetic analog La$_{3}$Cd$_{2}$As$_{6}$ are both semiconductors wherein a charge density wave phase transition takes place below 136~K and 279~K, respectively \cite{LaCe326}. The strong suppression of the charge density wave transition temperature ($T_{\rm CDW}$) from La$_{3}$Cd$_{2}$As$_{6}$ to Ce$_{3}$Cd$_{2}$As$_{6}$ is an indication that this phase may be completely suppressed by external pressure as Ce$_{3}$Cd$_{2}$As$_{6}$ presents a more compressed crystalline structure than La$_{3}$Cd$_{2}$As$_{6}$. Also, the activation energy decreases from 105(1)~meV to 74(1)~meV, going from La$_{3}$Cd$_{2}$As$_{6}$ to Ce$_{3}$Cd$_{2}$As$_{6}$ \cite{LaCe326}. By performing experiments under pressure on Ce$_{3}$Cd$_{2}$As$_{6}$ single crystals, one expects to completely suppress the CDW phase transition and to induce a metallic ground state.

To investigate this scenario, we report the electronic and magnetic properties of Ce$_{3}$Cd$_{2}$As$_{6}$ under pressure. At ambient pressure and low temperatures, an antiferromagnetic ordering of the Ce$^{3+}$ ions takes place below 4.0~K. The entropy recovered at T$_{\rm N}$ is about 75(5)~\% of the full doublet entropy $R\mathrm{ln}2$, which suggests the presence of either Kondo compensation or short-range interactions. The application of external pressure induces a metallic ground state in Ce$_{3}$Cd$_{2}$As$_{6}$ and completely suppresses the charge density wave order for pressures higher than 0.8(1)~GPa. However, superconductivity  above 2~K was not detected, possibly due to competition with the magnetically ordered state, as the antiferromagnetic transition temperature is enhanced by pressure, reaching 5.3~K at 3.8(1)~GPa. Finally, two different magnetic phases might be present in Ce$_{3}$Cd$_{2}$As$_{6}$ under pressure, due to the change in the resistivity anomaly associated to the antiferromagnetic phase transition.

\section{EXPERIMENTAL DETAILS}

Plate-like single crystals of Ce$_{3}$Cd$_{2}$As$_{6}$ and La$_{3}$Cd$_{2}$As$_{6}$ were grown by the vapor transport technique. First a polycrystalline seed of 1La/Ce:0.7Cd:2As was prepared via solid state reaction at 800~$^{\circ}$C. Then the polycrystalline powder was loaded along with iodine in a quartz tube, which was sealed in vacuum. The tube was kept in a temperature gradient from 830~$^{\circ}$C to 720~$^{\circ}$C. The initial polycrystalline material was kept in the hot zone, and the single crystals precipitated in the cold zone. The synthesized phase was checked by single crystal x-ray diffraction and energy-dispersive x-ray spectroscopy (EDX) resulting in 1:0.67:2 stoichiometry within experimental error. Magnetic susceptibility was measured in a commercial \textit{Quantum Design Magnetic Property Measurement System} (MPMS). The specific heat was measured using a commercial small mass calorimeter that employs a quasi-adiabatic thermal relaxation technique in a commercial \textit{Quantum Design Physical Property Measurement System} (PPMS). Electrical resistivity ($\rho$) was performed in a commercial PPMS along with AC bridges (Lakeshore AC372 Resistance Bridge or Linear Research LR-700 AC Resistance Bridge) or using the ACT option of the PPMS and measured in the $ab$-plane in a standard four-probe configuration. Pressures up to 1.9(1)~GPa were generated using a self-contained double-layer piston-cylinder-type Cu-Be pressure cell (PCC) with an inner-cylinder of hardened NiCrAl \cite{Nicklas}. For pressures to 3.8(1)~GPa a modified Bridgman anvil cell (mBAC) was used \cite{mBAC-Grenoble}. The pressure transmitting media used were Daphne oil 7373 and a 1:1 mixture of n-pentane and iso-pentane, respectively. Lead served as a low-temperature manometer in both pressure cells \cite{Nicklas}.
 
\begin{figure}[!tb]
\includegraphics[width=0.5\textwidth]{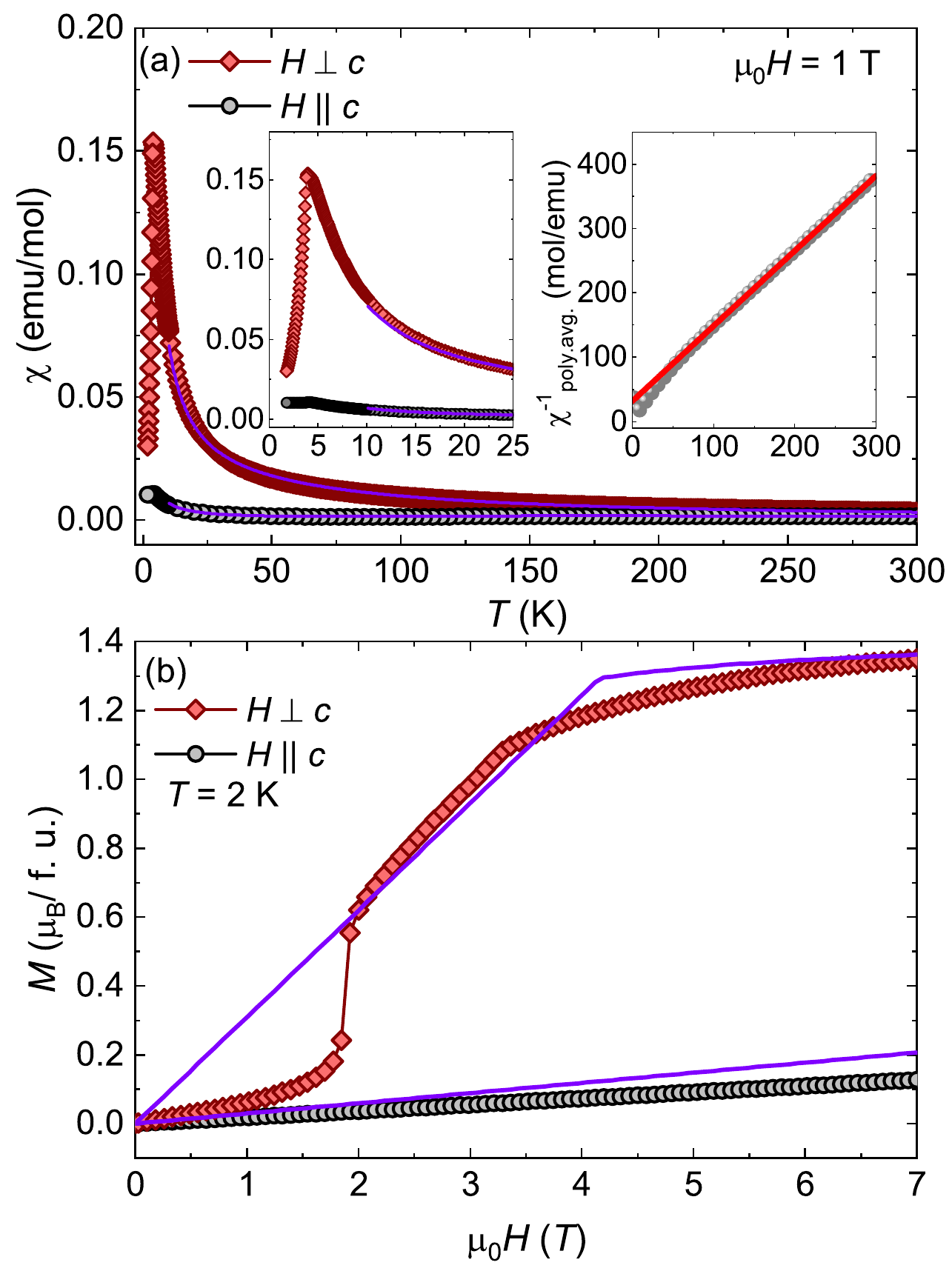}
\caption{(a) Magnetic susceptibility ($\chi = M/H$) as a function of temperature at $\mu_{0}H = 1$~T. The left inset presents a closer view of the magnetic susceptibility at low temperatures. The right inset shows the inverse of the polycrystalline average magnetic susceptibility as a function of temperature. The solid red line is an extrapolation of a linear fit performed at high temperatures ($T  \geq 150$~K). (b) Magnetization as a function of applied magnetic field at 2~K. The solid purple lines are fits using a CEF mean-field model.}
\label{chi}
\end{figure}

\begin{figure}[!htb]
\includegraphics[width=0.5\textwidth]{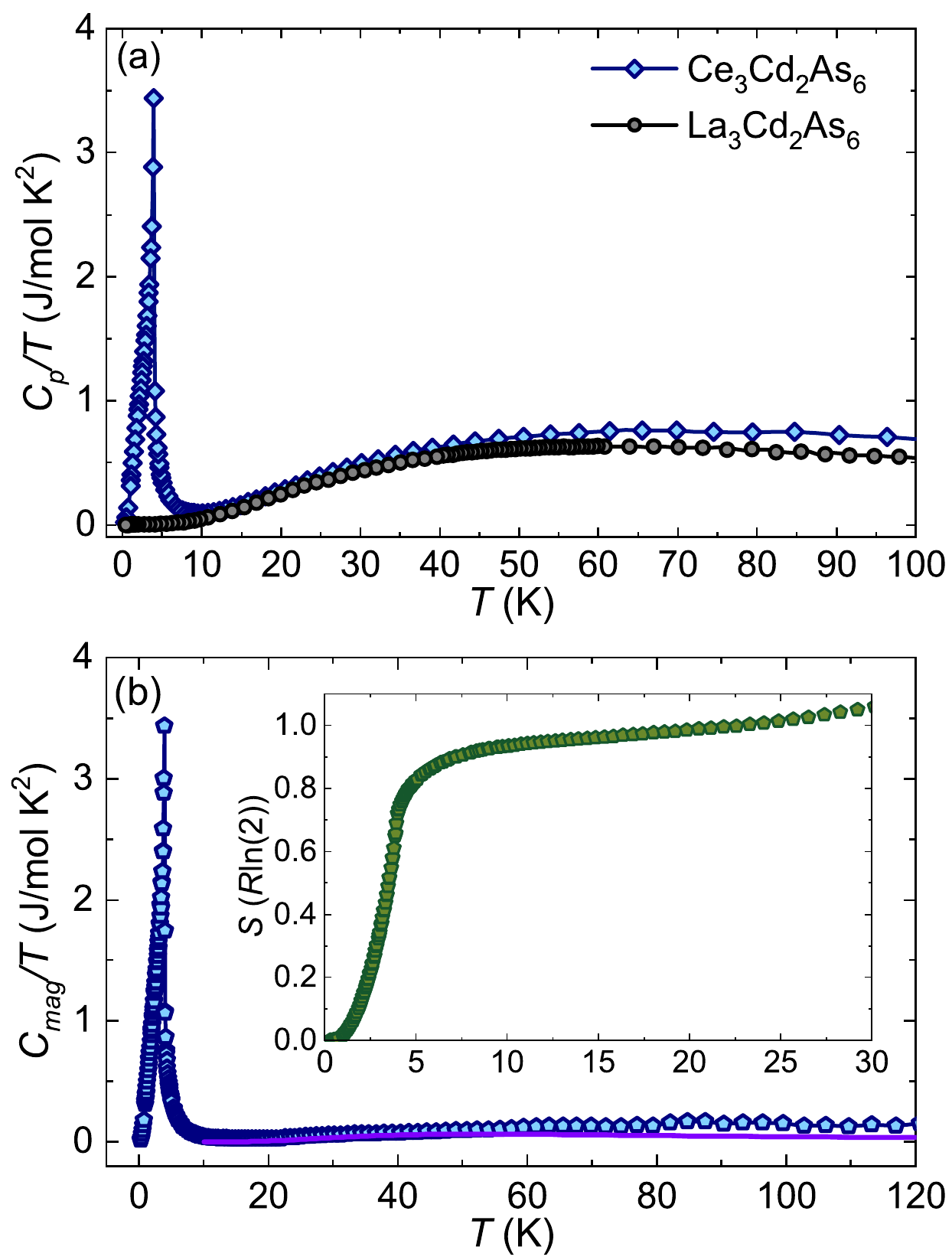}
\caption{(a) Specific heat measurements for Ce$_{3}$Cd$_{2}$As$_{6}$ and La$_{3}$Cd$_{2}$As$_{6}$. (b) Magnetic specific heat and entropy of Ce$_{3}$Cd$_{2}$As$_{6}$. The solid purple line is a fit using a CEF mean-field model. The inset shows the entropy obtained from the magnetic specific heat as a function of temperature.}
\label{cp}
\end{figure}

\section{RESULTS}

For simplicity, we refer to the tetragonal parent subcell $I4/mmm$ (Figure S1) as the structure of Ce$_{3}$Cd$_{2}$As$_{6}$ in further discussions, without affecting the interpretation of our results. Figure~\ref{chi}(a) presents the magnetic susceptibility ($\chi = M/H$) as a function of temperature at $\mu_{0}H = 1$~T applied parallel and perpendicular to the $c$-axis of Ce$_{3}$Cd$_{2}$As$_{6}$. These measurements find that the $c$-axis is the hard axis. At low temperatures, the Ce$^{3+}$ moments order antiferromagnetically below $T_{\rm N} = 4.0$~K, as shown in the left inset of Fig.~\ref{chi}(a). The right inset shows the inverse of the magnetic susceptibility for the polycrystalline average ($\frac{1}{3}\chi_{//c} + \frac{2}{3}\chi_{\perp c}$). By performing linear fits in the high-temperature range ($T  \geq 150$~K), a Ce$^{3+}$ effective moment of $2.6(1)$~$\mu_{B}$ was obtained, in agreement with the theoretical value for a free Ce$^{3+}$ ion. From the Curie-Weiss fit, we also extracted a Weiss temperature of $-27(1)$~K. Figure~\ref{chi}(b) displays the magnetization ($M$) behavior as a function of applied magnetic field at 2~K. A metamagnetic transition is observed at 1.9~T when fields are applied perpendicular to $c$-axis. Further, a saturation starts to develop at 3.4~T. For fields parallel to the hard $c$-axis, the magnetization linearly increases with increasing magnetic field. 

Figure~\ref{cp}(a) displays the specific heat divided by temperature ($C_{p}/T$) per mole of lanthanide for both Ce$_{3}$Cd$_{2}$As$_{6}$ and La$_{3}$Cd$_{2}$As$_{6}$ as a function of temperature. A lambda type peak defines the antiferromagnetic transition temperature at $T_{\rm N} = 4.0$~K, in agreement with susceptibility measurements. By subtracting the phonon contribution given by the specific heat of its non-magnetic analog La$_{3}$Cd$_{2}$As$_{6}$, we obtain the magnetic specific heat ($C_{mag}/T$), displayed in Fig.~\ref{cp}(b) from 0.33~K to 30~K. We note that due to the small mass of the single crystals, specific heat measurements of La$_{3}$Cd$_{2}$As$_{6}$ were collected on a mosaic of 10 single crystals of total mass around 1~mg, which could lead to errors in the phonon subtraction. The bump centered around 90~K may be associated with a Schottky anomaly due to first crystal field excited state.

\begin{figure}[!t]
	\includegraphics[width=0.41\textwidth]{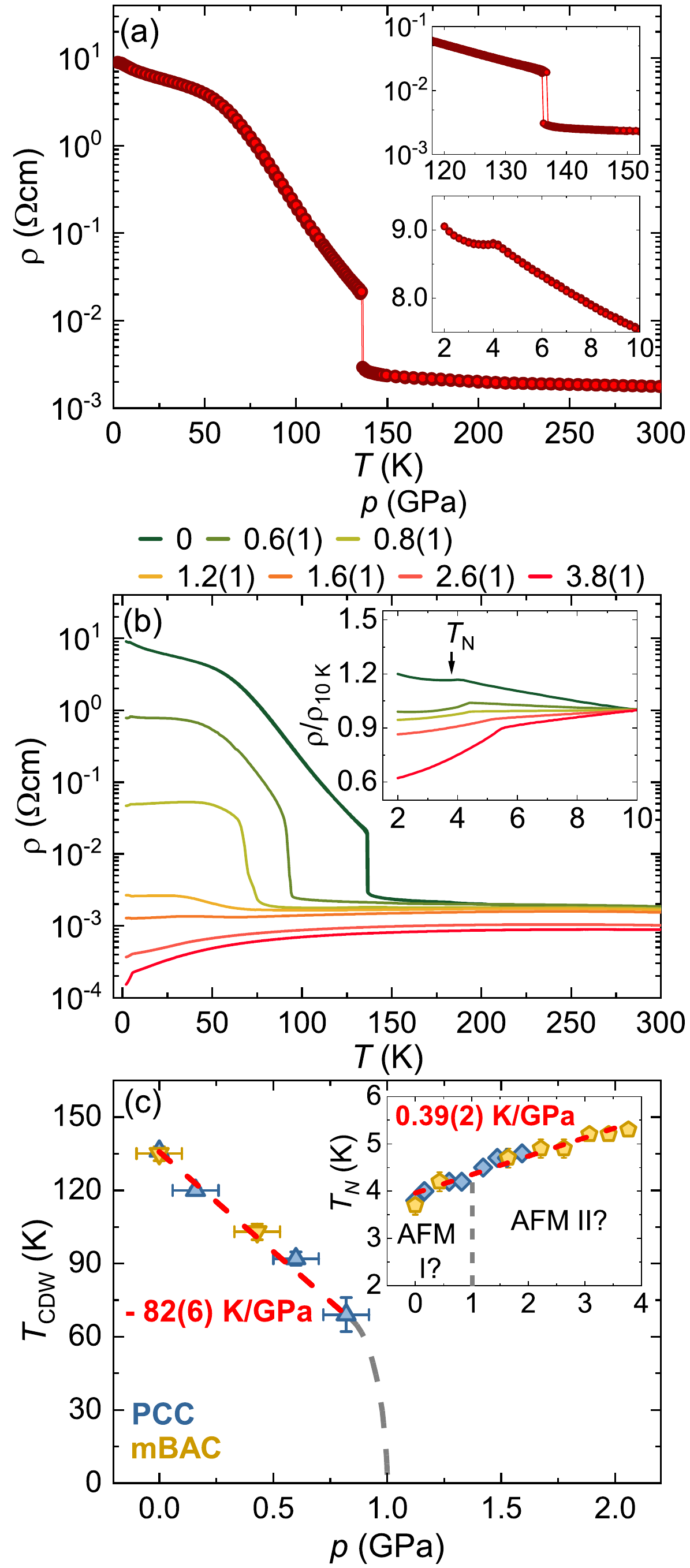}
	\caption{(a) Electrical resistivity as a function of temperature for Ce$_{3}$Cd$_{2}$As$_{6}$. The top inset shows a closer view to the charge density wave transition at high temperatures. The bottom inset displays a zoomed-in view of the low-temperature behavior. (b) Electrical resistivity as a function of temperature at several pressures. The data up to 1.6(1)~GPa were taken using a PCC. The 2.6(1)~GPa and 3.8(1)~GPa measurements were performed in a mBAC. We note that resistivity values for the mBAC results were obtained using the  expected resistivity value for Ce$_{3}$Cd$_{2}$As$_{6}$ at room temperature. The inset displays the low temperature range ($T \leq 10$~K) of the electrical resistivity normalized by its 10~K value. The arrow points to $T_{\rm N}$. (c) Temperature-pressure phase diagram for Ce$_{3}$Cd$_{2}$As$_{6}$. The dashed red lines are linear fits and the dashed gray lines are a guide to the eye.}
	\label{rhopress}
\end{figure}

The solid purple lines in Figs.~\ref{chi} and in Fig.~\ref{cp}(b) are fits using a crystalline electric field (CEF) mean field model considering competing nearest-neighbor interactions and the tetragonal CEF Hamiltonian: $\mathcal{H} = g_{J}\mu_{B}\boldsymbol{H}\cdot \boldsymbol{J} + z_{i}J_{i}^{ex} \cdot  \langle J^{ex}\rangle + B^{0}_{2}O^{0}_{2} + B^{0}_{4}O^{0}_{4} + B^{4}_{4}O^{4}_{4}$, where $g_{J}$ is the Land\'e $g$-factor, $\mu_{B}$ is the Bohr magneton, $\boldsymbol{H}$ is the applied magnetic field and $\boldsymbol{J}$ is the total angular momentum. $z_{i}J_{i}^{ex}$ represents the $J_{i}$ mean field exchange interactions ($i = $AFM, FM) between the $z_{i}$ nearest neighbors. $B^{m}_{n}$ are the CEF parameters and the $O^{m}_{n}$ are the Steven's operators  \cite{pagliuso2006}. By simultaneously performing fits to $\chi(T)$,  $M(H)$ and $C_{mag}/T$ data, we extract the CEF scheme and two exchange parameters for this compound. For the CEF parameters we obtained the following values: $B^{0}_{2} \approx 22.8$~K, $B^{0}_{4} \approx - 0.3$~K and $B^{4}_{4} \approx -4.3$~K. These parameters imply a ground state composed of a $\Gamma_{6} = |\pm1/2\rangle$ doublet, a first excited state $\Gamma_{7}^{2}= 0.40 |\pm5/2\rangle + 0.91 |\mp3/2\rangle$ doublet at 160~K and a second excited state of a $\Gamma_{7}^{1} = 0.91 |\pm5/2\rangle - 0.40 |\mp3/2\rangle$ doublet at 480~K. Interestingly, in spite of its $\Gamma_{6}$ ground state, Ce$_{3}$Cd$_{2}$As$_{6}$  is antiferromagnetic and does not follow the general trend of the ``112'' family of compounds, possibly due to its body-centered $I4/mmm$ subcell and superstructure modification. For the exchange parameters we obtain $z_{AFM}J_{AFM}^{ex} = 0.8$~K and $z_{FM}J_{FM}^{ex} = -1.3$~K. The presence of ferromagnetic and antiferromagnetic exchange interactions is required to simulate the field-induced transition when $H || ab$. We note that the CEF parameters acquired from the fits of macroscopic data may not be fully accurate and must be treated with caution. The fact that this simple model does not account for the sharp feature around 1.9~T in the magnetization for $H || ab$ indicates that more exchange parameters, or anisotropic terms such as $J_x$, $J_y$ and $J_z$ may be needed to fully fit the macroscopic data of this material. Nevertheless, this model yields important information about Ce$_{3}$Cd$_{2}$As$_{6}$, as the $\Gamma_{6}$ ground state and that the Ce$^{3+}$ ions are much localized. The integral of $C_{mag}/T$ vs $T$ gives the magnetic entropy presented in the inset of Fig.~\ref{cp}(b) that reaches 75(5)~\% of the doublet entropy $R\textrm{ln}(2)$ at $T_{\rm N}$. A partial Kondo effect or short range magnetic correlations could be responsible for the presence of entropy above $T_{\rm N}$. 

Figure~\ref{rhopress}(a) displays the electrical resistivity ($\rho$) of Ce$_{3}$Cd$_{2}$As$_{6}$, which reveals semiconducting behavior. At 136~K, a sharp increase of the resistivity (displayed in the top inset of Fig.~\ref{rhopress}(a)) might be associated to a charge density wave transition. Ce$_{3}$Cd$_{2}$As$_{6}$ presents Cd vacancies ordered in a stripe pattern, that breaks the parent compounds’ fourfold symmetry and are accompanied by several structural modifications \cite{LaCe326}. The displacement of Cd ions also affects their ligands (As1, As2, and As3), leading to distorted As square nets \cite{LaCe326}. These As square nets resemble the Te nets that host similar superstructures to Ce$_{3}$Cd$_{2}$As$_{6}$ and also present CDW phases, such as K$_{0.33}$Ba$_{0.67}$AgTe$_{2}$ \cite{KBaAgTe2}. Around 60~K, an inflection in electrical resistivity may indicate the presence of additional conduction channels prone to exist at low temperatures in narrow-gap semiconductors due to disorder \cite{Book2}. At low temperatures, there is a well-defined kink around 4~K that reflects the loss of spin-disorder scattering below the antiferromagnetic transition, as displayed in the bottom inset of Fig.~\ref{rhopress}(a). The transition temperature values for the CDW and AFM transitions were inferred from the temperature dependent resistivity data by identifying the local extrema in $d(\textrm{ln}\rho)/dT$ \cite{CDW_rhodef} and $d\rho/dT$ \cite{Fisher_rho} data respectively. For $T_{\rm CDW}$, the error bars were estimated based on Gaussian fits of the peaks  observed in the temperature derivative of the natural logarithm of the resistivity and for $T_{\rm N}$, the broadness of the anomalies were used to estimate the error bars. More details about the criteria used for the transition temperatures can be found in the Supplemental Material \cite{SM}. 

\begin{figure}[!t]
	\includegraphics[width=0.43\textwidth]{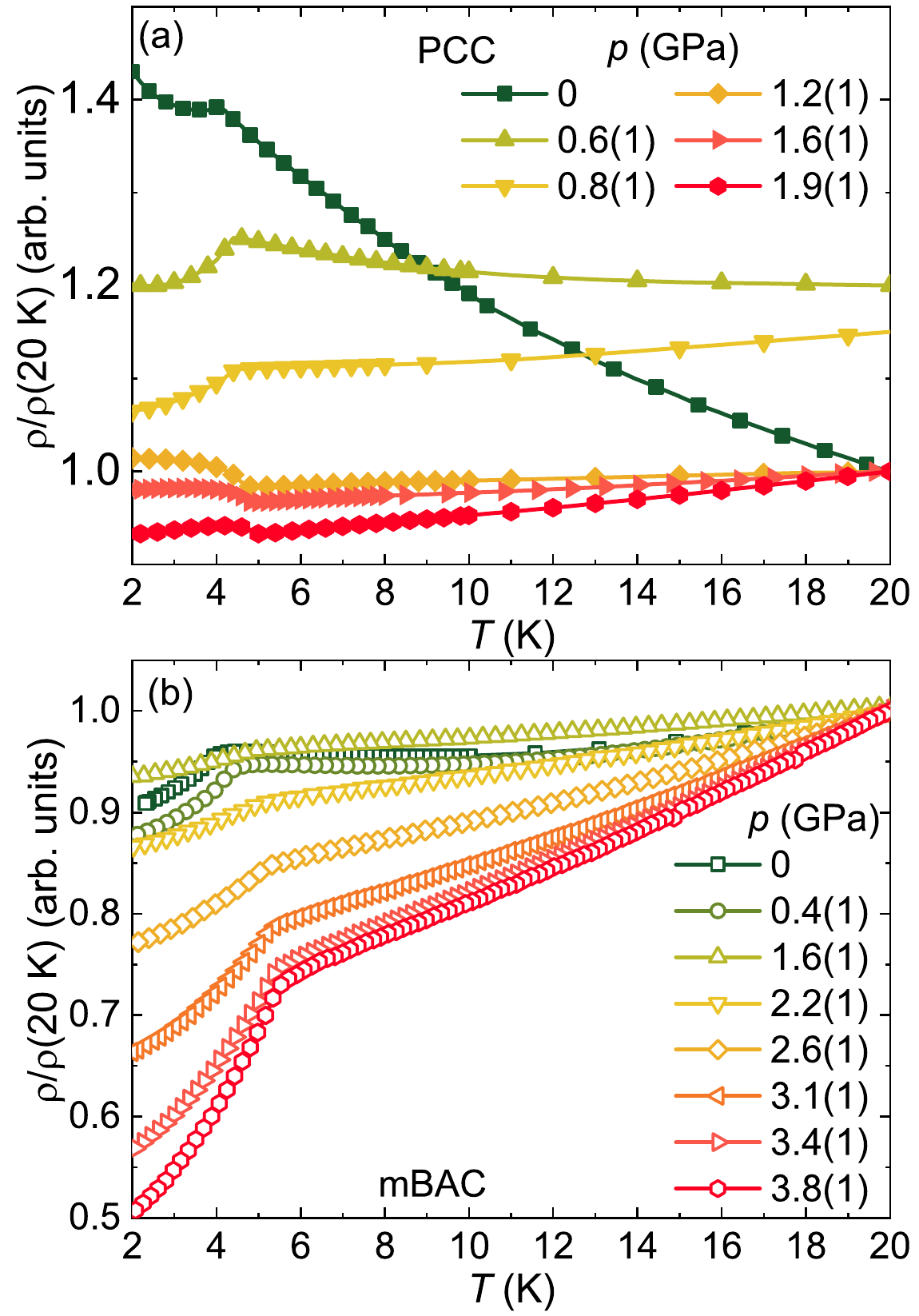}
	\caption{Normalized electrical resistivity as a function of temperature for Ce$_{3}$Cd$_{2}$As$_{6}$ for measurements using a PCC (a) and a mBAC (b) at selected pressures. The curves at 0.6(1)~GPa and 0.8(1)~GPa in panel (a) were vertically shifted to improve visualization.}
	\label{scatteringcomp}
\end{figure}  

Figure~\ref{rhopress}(b) displays the electrical resistivity as a function of temperature at several pressures. Overall the application of external pressure suppresses the semiconducting behavior and induces a metallic ground state, i.e. a positive $d\rho/dT$, as pressures increase above $\approx 2$~GPa. Further increasing pressure enhances the metallic behavior of the material, which presents a RRR $ \approx $ 6 and a room temperature resistivity of the order of 0.9~m$\Omega$cm at 3.8(1)~GPa. Furthermore, external pressure strongly suppresses the charge-density wave transition temperature ($T_{\rm CDW}$) at a initial rate of $-82(6)$~K/GPa, which cannot be identified anymore above 0.8(1)~GPa (see Fig.~\ref{rhopress}(c) and Fig.~2 of the Supplemental Material \cite{SM}). We note that the critical pressure of suppression of the CDW order cannot be unambiguously determined by resistivity measurements alone. Therefore, the dashed gray lines in Fig.~\ref{rhopress}(c) represent only one possibility for the critical pressure. A suppression of the CDW order with increasing pressure was expected, as discussed before, because external pressure can destroy the nesting vectors responsible for the CDW phase, closing the energy gaps created by this phase and favoring a metallic ground state \cite{4HbTaS2}. For Ce$_{3}$Cd$_{2}$As$_{6}$, the nesting vector responsible for the putative CDW phase is unknown, therefore electronic band structure calculations and angle resolved photoemission spectroscopy (ARPES) are in need to investigate the Fermi surface of this material. A metallic state was observed at high pressures in many semiconducting materials in which the gap is created by CDW order, e.g. CeRhAs \cite{CeRhAs,CeRhAspress1,CeRhAspress2}, Sr$_{3}$Ir$_{4}$Sn$_{13}$, Ca$_{3}$Ir$_{4}$Sn$_{13}$ \cite{CDWSC}, La$_{3}$Co$_{4}$Sn$_{13}$ \cite{La3413-pressure} and many transition metal dichalcogenides \cite{TMDS1,TMDS2,TMDS3,TMDS4}. This behavior is in contrast with the observed gap evolution as a function of pressure for Kondo insulators such as Ce$_{3}$Bi$_{4}$Pt$_{3}$ wherein the gap is created by the hybridization between the conduction electrons and 4$f$ electrons. Applying external pressure enhances this hybridization, leading to larger energy gaps at high pressures \cite{CeRhAspress1,pressureCe343}. Nevertheless, x-ray diffraction experiments under pressure are desired to investigate the lattice evolution as a function of pressure. 

Conversely, the antiferromagnetic transition temperature is gradually increased by increasing pressure, reaching a $T_{\rm N}$ of 5.3~K at 3.8(1)~GPa, as shown in the inset of Fig.~\ref{rhopress}(c). The enhancement of $T_{\rm N}$ can be associated with two intertwined factors. First, the application of external pressure leads to an increase in the density of states at the Fermi level, as the material becomes more metallic with increasing pressure, favoring higher transition temperatures. Second, the exchange interaction between the localized moments might also be enhanced by pressure, favoring the magnetic order. This behavior indicates that Ce$_{3}$Cd$_{2}$As$_{6}$ is in the localized limit of the Doniach diagram \cite{Doniach} and that experiments at higher pressures are in need to elucidate if the Kondo effect can be tuned to overcome the antiferromagnetism, suppressing $T_{\rm N}$ and maybe inducing a quantum critical point. We note, that the exchange interaction responsible for the magnetism in Ce$_{3}$Cd$_{2}$As$_{6}$ is still an open question. Figure~\ref{rhopress}(c) summarizes the evolution of the transition temperatures of Ce$_{3}$Cd$_{2}$As$_{6}$ as a function of pressure using both the piston cylinder cell (PCC) and the Bridgman cell (mBAC). A detailed summary of the results obtained with both cells is presented in the Supplemental Material \cite{SM}. 

Notably, under pressure the signature associated with magnetic order changes at $T_{\rm N}$ in the PCC measurements, as shown in Figure~\ref{scatteringcomp}(a). At first, a sudden decrease in the resistivity below $T_{\rm N}$ is observed due to the loss of spin disorder scattering. However, at pressures of 1.2(1)~GPa and higher, this feature changes to an upturn at $T_{\rm N}$ possibly due to the formation of a superzone gap. This change is not observed in the mBAC measurements (Figure~\ref{scatteringcomp}(b)), in which the antiferromagnetic order lead to a sudden decrease in the resistivity for all studied pressures. We speculate that this difference is caused by different current directions. Although the currents in both pressure cells were in the $ab$-plane, fortunately, the current direction in the PCC allowed us to detect this change of scattering. Moreover, this difference indicates that two different magnetic phases might be present under pressure, which seems to be related with the suppression of the CDW phase in Ce$_{3}$Cd$_{2}$As$_{6}$ as shown in Fig.~\ref{rhopress}(c). Another possible scenario is a change in the Fermi surface as the CDW is completely suppressed without any changes in the magnetic order. Nevertheless, the determination of the origin of this scattering type change requires more experiments with well defined current directions and is beyond the scope of this manuscript.

\section{CONCLUSIONS}

In summary, we report the electronic and magnetic properties of the newly-discovered compound Ce$_{3}$Cd$_{2}$As$_{6}$ \cite{LaCe326}. At ambient pressure, Ce$_{3}$Cd$_{2}$As$_{6}$ is a narrow-gap semiconductor and presents a charge density wave phase transition at 136~K. Specific heat measurements show a lambda type peak at 4.0~K associated with the antiferromagnetic ordering of the Ce$^{3+}$ ions in Ce$_{3}$Cd$_{2}$As$_{6}$. Furthermore, CEF fits of $\chi(T)$,  $M(H)$ and $C_{mag}/T$ yielded a CEF scheme with the following parameters: $B^{0}_{2} \approx 22.8$~K, $B^{0}_{4} \approx - 0.3$~K and $B^{4}_{4} \approx -4.3$~K. These parameters result in a ground state composed of a $\Gamma_{6} = |\pm1/2\rangle$ doublet. Moreover, two competing exchange parameters were estimated with the following values $z_{AFM}J_{AFM}^{ex} = 0.8$~K and $z_{FM}J_{FM}^{ex} = -1.3$~K. 

The application of external pressure on Ce$_{3}$Cd$_{2}$As$_{6}$ single crystals induced a metallic ground state and strongly suppressed the charge density wave phase, which was not observed for pressures above 0.8(1)~GPa. Coincident with the suppression of the CDW phase, a change of the anomaly in the resistivity associated to the antiferromagnetic order was observed with increasing pressure, indicating that Ce$_{3}$Cd$_{2}$As$_{6}$ might host two different antiferromagnetic orders under pressure.  The details of the ordering wave vectors of these phase will have to be studied in future. The antiferromagnetic state was favored by pressure as the system became metallic, reaching a $T_{\rm N}$ of 5.3 K at 3.8(1)~GPa. Experiments at higher pressures are in need to investigate if $T_{\rm N}$ can be suppressed maybe leading to a quantum critical point.

\begin{acknowledgments}

We would like to acknowledge G. S. Freitas, M. C. Rahn and P. G. Pagliuso for fruitful discussions. This project has received funding from the European Union’s Horizon 2020 research and innovation programme under the Marie Sk\l{}odowska-Curie grant agreement No 101019024. Work at Los Alamos was performed under the auspices of the U.S. Department of Energy, Office of Basic Energy Sciences, Division of Materials Science and Engineering.  M.M.P. also acknowledges support from the Sao Paulo Research Foundation (FAPESP) grants 2015/15665-3, 2017/25269-3, 2017/10581-1, CAPES and CNPq, Brazil. Scanning electron microscope and energy dispersive X-ray measurements were performed at the Center for Integrated Nanotechnologies, an Office of Science User Facility operated for the U.S. Department of Energy (DOE) Office of Science. Work at Ames Laboratory (L.X., S.L.B., and P.C.C.) was supported by the U.S. Department of Energy, Office of Science, Basic Energy Sciences, Materials Sciences and Engineering Division. Ames Laboratory is operated for the U.S. Department of Energy by Iowa State University under Contract No. DE-AC02-07CH11358. L.X. and R.A.R. were supported, in part, by the Gordon and Betty Moore Foundation’s EPiQS Initiative through Grant No. GBMF4411 and by the W. M. Keck Foundation. 

\end{acknowledgments}

\bibliography{basename of .bib file}

\begin{figure*}[!t]
	
	\centering
	\includegraphics[width=1\textwidth]{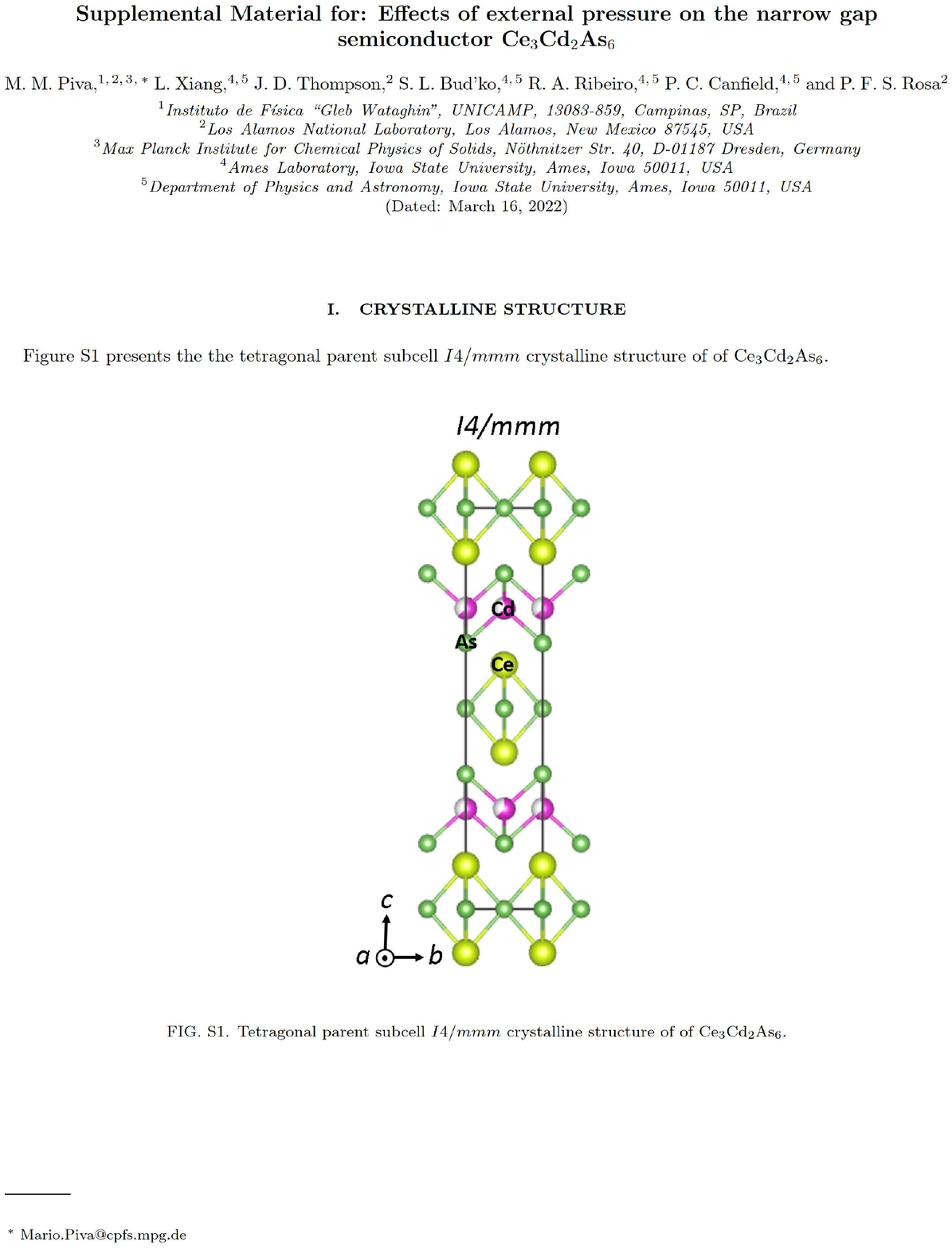}
\end{figure*}

\begin{figure*}[!t]
	
	\centering
	\includegraphics[width=1\textwidth]{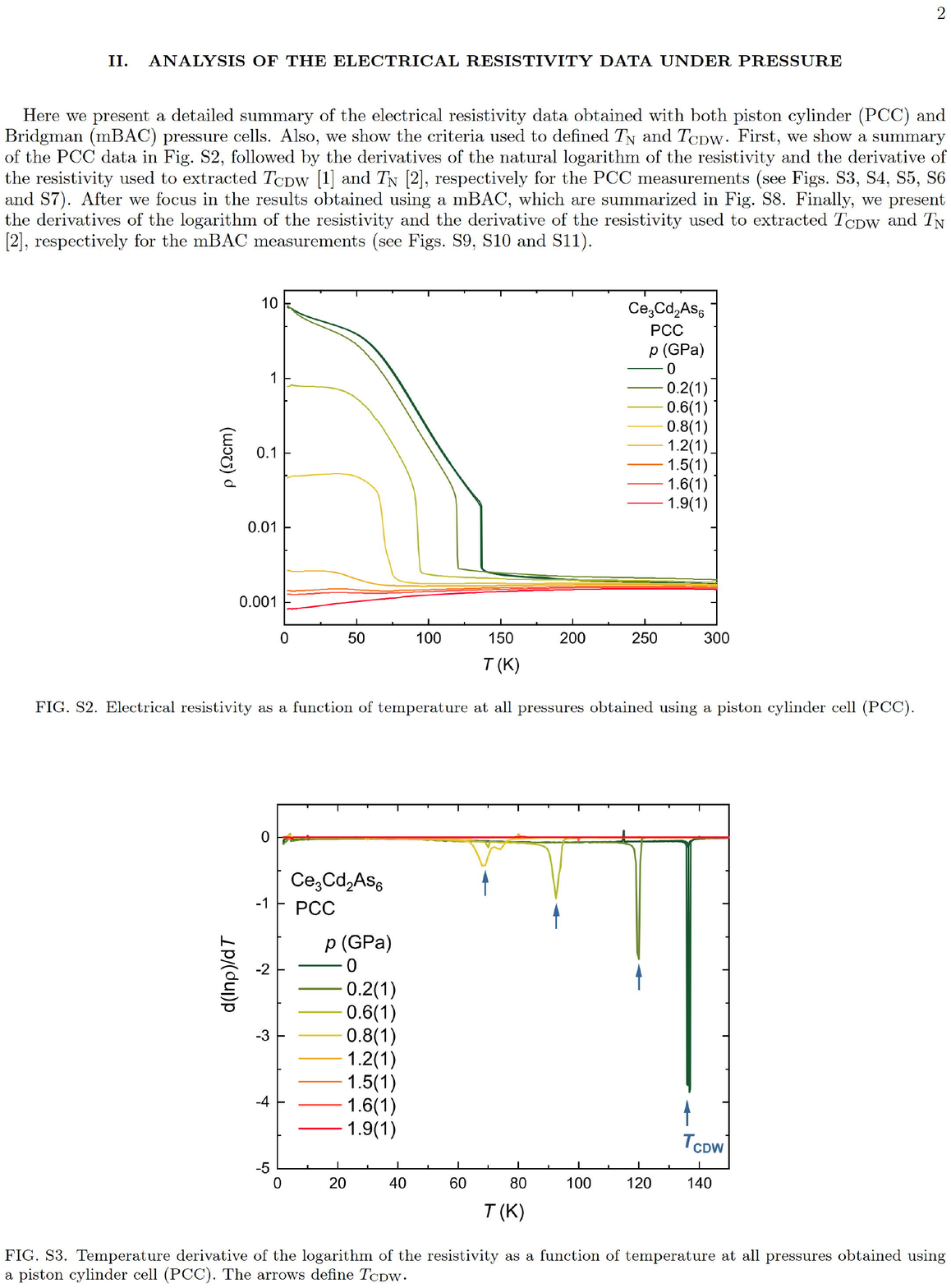}
\end{figure*}

\begin{figure*}[!t]
	
	\centering
	\includegraphics[width=1\textwidth]{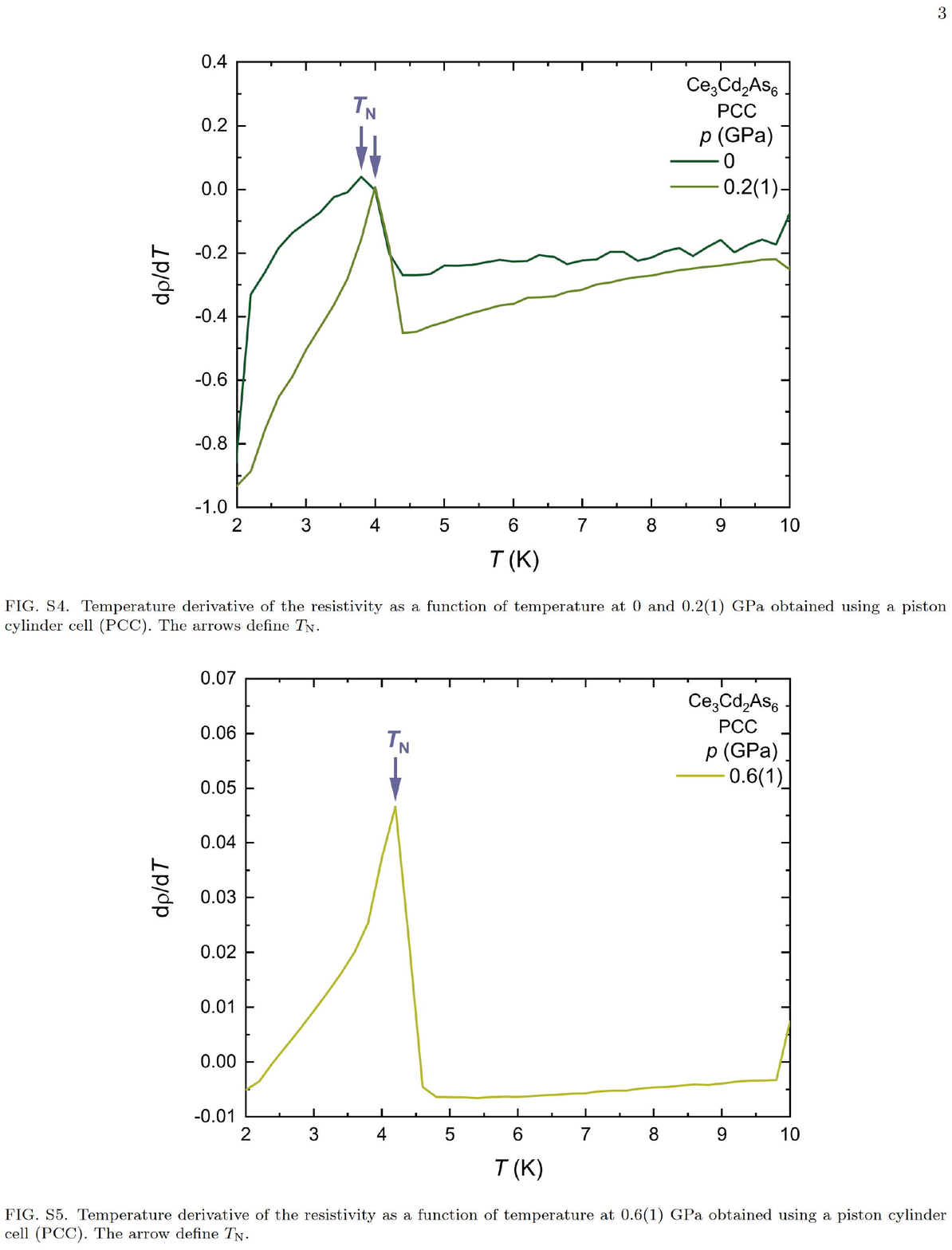}
\end{figure*}

\begin{figure*}[!t]
	
	\centering
	\includegraphics[width=1\textwidth]{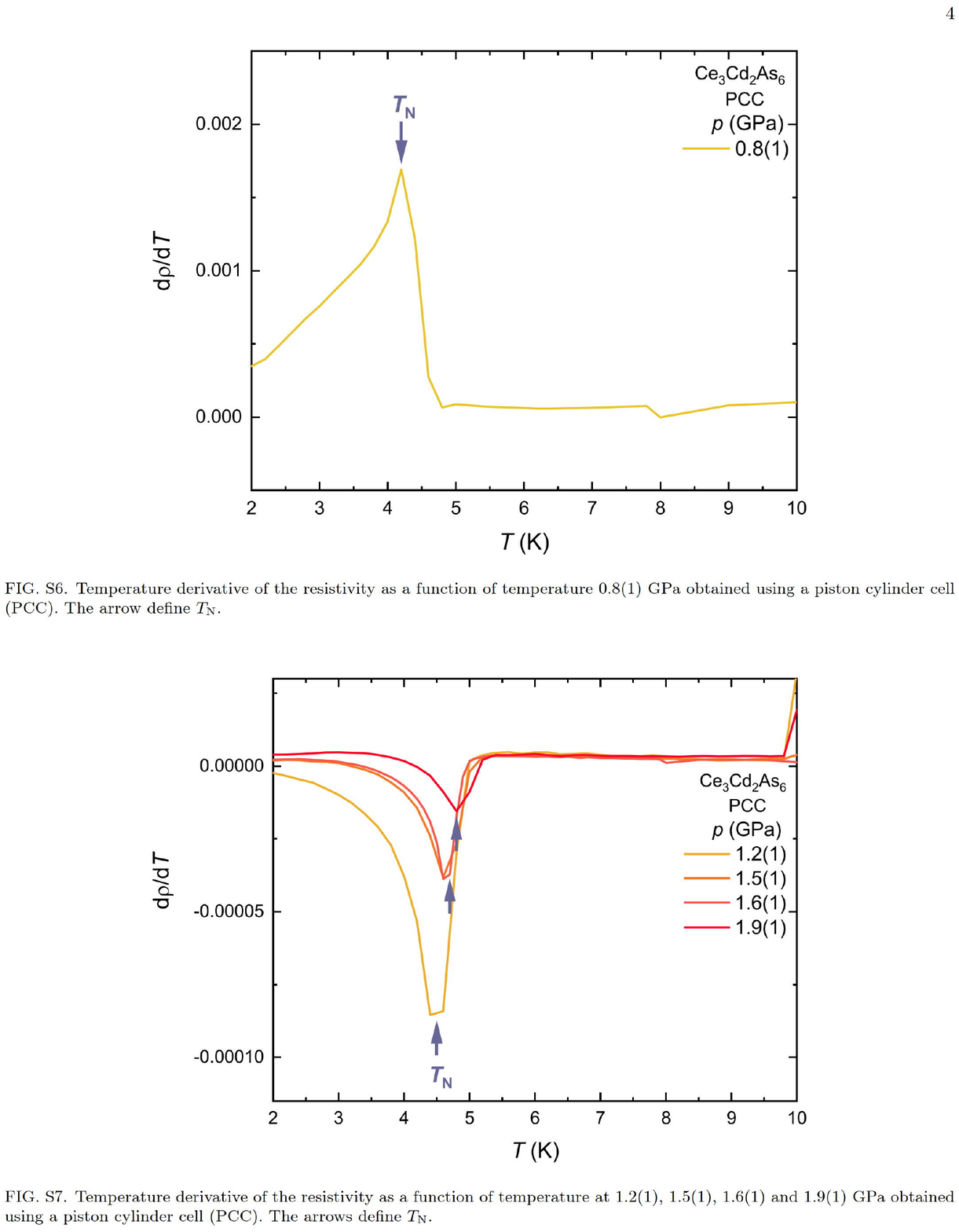}
\end{figure*}

\begin{figure*}[!t]
	
	\centering
	\includegraphics[width=1\textwidth]{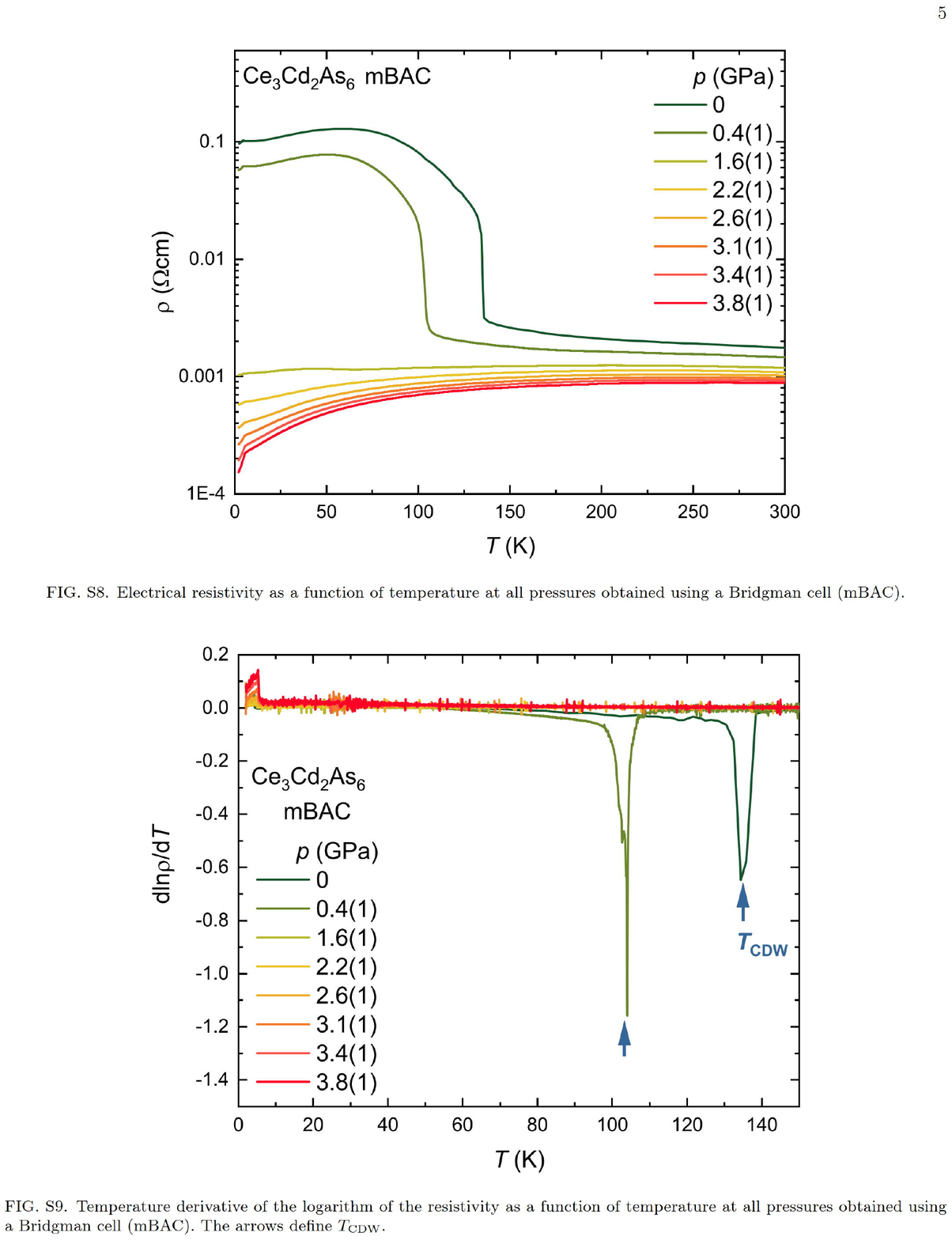}
\end{figure*}

\begin{figure*}[!t]
	
	\centering
	\includegraphics[width=1\textwidth]{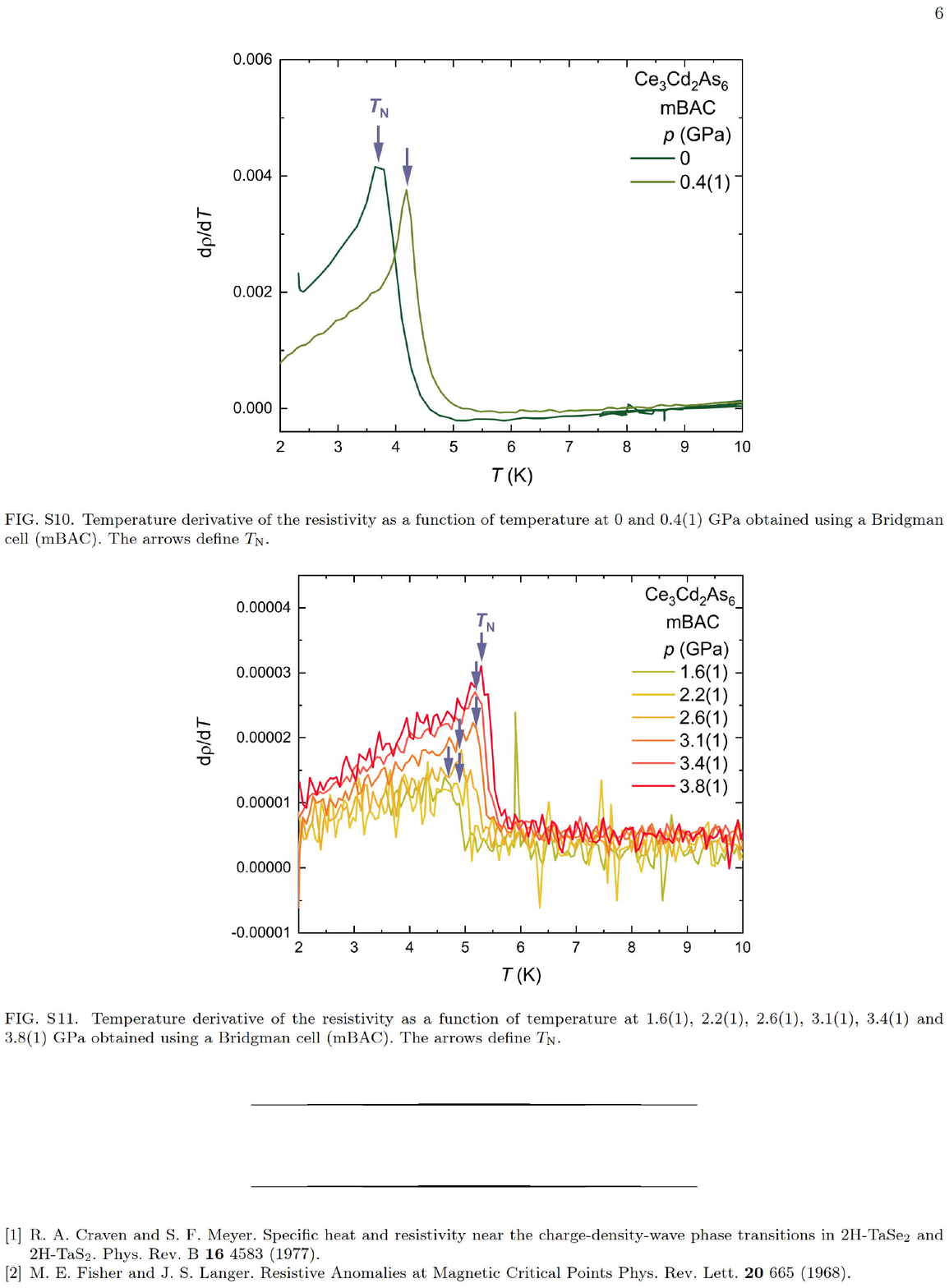}
\end{figure*}

\end{document}